\documentclass[ aps,%
floatfix,%
final,%
notitlepage,%
oneside,%
onecolumn,%
nobibnotes,%
nofootinbib,%
superscriptaddress,%
showpacs,%
]%
{revtex4}

\usepackage{epsfig}
\usepackage{amsfonts}
\usepackage{amsmath}
\usepackage{epsfig}
\usepackage{graphics}

\newcommand{\beq}{\begin{eqnarray}}\newcommand{\eeq}{\end{eqnarray}}
\newcommand{\beqa}{\begin{eqnarray*}}\newcommand{\eeqa}{\end{eqnarray*}}

\begin{document}

\title{Leading twist distribution amplitudes of $P$-wave nonrelativistic mesons. }
\author{V.V. Braguta}

\email{braguta@mail.ru}

\author{A.K. Likhoded}

\email{Likhoded@ihep.ru}

\author{A.V. Luchinsky}

\email{Alexey.Luchinsky@ihep.ru}

\affiliation{Institute for High Energy Physics, Protvino, Russia}

\begin{abstract}
This paper is devoted to the study of the leading twist distribution amplitudes of $P$-wave nonrelativistic 
mesons. It is shown that at the leading order approximation in relative velocity of quark-antiquark pair 
inside the mesons these distribution amplitudes can be expressed through one universal function. As an example, 
the distribution amplitudes of $P$-wave charmonia mesons are considered. Within QCD sum rules the model for the 
universal function of $P$-wave charmonia mesons is built. In addition, it is found the relations between 
the moments of the universal function and the nonrelativistic QCD matrix elements 
that control relativistic corrections to any amplitude involving $P$-wave charmonia. 
Our calculation shows that characteristic size of these corrections is of order of $\sim 30 \%$.

\end{abstract}


\maketitle

\newcommand{\ins}[1]{\underline{#1}}
\newcommand{\subs}[2]{\underline{#2}}
\vspace*{-1.cm}
\section{Introduction}

Hard exclusive processes are very interesting both from theoretical and experimental points of view. 
Commonly, theoretical approach to the description of such processes is based on 
the factorization theorem \cite{Lepage:1980fj, Chernyak:1983ej}. Within this theorem the amplitude of hard exclusive process
can be separated into two parts. The first part is partons production at very small 
distances, which can be treated within perturbative QCD. The second part is 
hardronization of the partons at larger distances. This part contains  information about 
nonperturbative dynamic of strong interaction. For hard exclusive processes it can be 
parameterized by process independent distribution amplitudes (DA), which can be considered as hadrons' wave functions
at light like separation between the partons in the hadron. It should stressed that 
DAs are very important for the calculation of the amplitude of any hard exclusive
process. 

Recently, the leading twist DAs of $S$-wave nonrelativistic mesons have become the object of intensive 
study
\cite{Bodwin:2006dm, Ma:2006hc, Braguta:2006wr, Braguta:2007fh, Braguta:2007tq, Choi:2007ze, Feldmann:2007id, Bell:2008er}.
Knowledge about these DAs allowed one to build some models for $S$-wave charmonia DAs, 
that can be used in practical calculations. In this paper  general properties of the 
leading twist DAs of $P$-wave nonrelativistc mesons will be studied. The results of this 
study will be used to build the model for the DAs of $P$-wave charmonia mesons, that 
can be used in calculations. 

This paper is organized as follows. In the next section all definitions of the DAs of $P$-wave
mesons will be given. These DAs will be studied in section III at the leading order approximation 
in relative velocity of quark-antiquark pair inside $P$-wave meson. In section IV QCD sum rules 
will be applied to the calculation of the moments of the $P$-wave charmonia DAs. Using the results 
of this study the model of the $P$-wave charmonia DAs will be build in section V. In the last section
the results of this paper will be summarized.

\section{Definitions of the distribution amplitudes.}

In this section the definitions of the leading twist distribution amplitudes (DA) of $P$-wave nonrelativistic mesons
will be given. In the conventional quark model nonrelativistic mesons are quark-antiquark($Q \bar Q$) 
bound states. In these mesons the quark-antiquark pair can be in the spin singlet or spin triplet states. 
Since orbital momentum of the quark-antiquark pair is unity one can conclude that there are  four $P$-wave
mesons: $\chi_0 (^3P_0), \chi_1(^3P_1), \chi_2(^3P_2), h(^1P_1)$. The leading twist DAs of 
these mesons can be defined as follows. \\
{\bf for the $\chi_0$ meson:}
\beq
{\langle \chi_0(p) | {\bar Q} (z) \gamma_{\mu}  [z,-z] Q(-z) |0  \rangle} &=&
f_{\chi_0} p_{\mu} \int^1_{-1} d \xi \,e^{i(pz) \xi}
\phi_{\chi_0} ( \xi, \mu),
\label{chi0}
\eeq
{\bf for the $\chi_1$  meson:}
\beq
{\langle \chi_1(p, \epsilon_{\lambda=0}) | {\bar Q} (z) \gamma_{\mu} \gamma_5  [z,-z] Q(-z) |0  \rangle} &=&
f'_{\chi_1} p_{\mu} \int^1_{-1} d \xi \,e^{i(pz) \xi}
\phi'_{\chi_1} ( \xi, \mu), \nonumber \\
{\langle \chi_1(p, \epsilon_{\lambda=\pm 1}) | {\bar Q} (z) \sigma_{\mu \nu}  [z,-z] Q(-z) |0  \rangle} &=&
f_{\chi_1} e_{\mu \nu \alpha \beta} \epsilon^{\alpha} p^{\beta}  \int^1_{-1} d \xi \,e^{i(pz) \xi}
\phi_{\chi_1} ( \xi, \mu), 
\label{chi1}
\eeq
{\bf for the $h$  meson:}
\beq
{\langle h(p, \epsilon_{\lambda=0}) | {\bar Q} (z) \gamma_{\mu} \gamma_5  [z,-z] Q(-z) |0  \rangle} &=&
f_{h} p_{\mu} \int^1_{-1} d \xi \,e^{i(pz) \xi}
\phi_{h} ( \xi, \mu), \nonumber \\
{\langle h(p, \epsilon_{\lambda=\pm 1}) | {\bar Q} (z) \sigma_{\mu \nu}  [z,-z] Q(-z) | 0 \rangle} &=&
f'_{h} e_{\mu \nu \alpha \beta} \epsilon^{\alpha} p^{\beta}  \int^1_{-1} d \xi \,e^{i(pz) \xi}
\phi'_{h} ( \xi, \mu), 
\label{h}
\eeq
{\bf for the $\chi_{2}$  meson:}
\beq
{\langle \chi_2(p, \epsilon_{\lambda=0}) | {\bar Q} (z) \gamma_{\mu}   [z,-z] Q(-z) | 0 \rangle} &=&
f_{\chi_2} p_{\mu} \int^1_{-1} d \xi \,e^{i(pz) \xi}
\phi_{\chi_2} ( \xi, \mu), \nonumber \\
{\langle \chi_2(p, \epsilon_{\lambda=\pm 1}) | {\bar Q} (z) \sigma_{\mu \nu}  [z,-z] Q(-z) | 0 \rangle} &=&
\tilde {f}_{\chi_2} M_{\chi_2}  
(  \rho_{\mu}  p_{\nu} - \rho_{\nu}  p_{\mu} )  \int^1_{-1} d \xi \,e^{i(pz) \xi}
\tilde {\phi}_{\chi_2} ( \xi, \mu),~~~ \rho_{\mu}= \frac {\epsilon_{\mu \sigma} z^{\sigma}} {pz}, 
\label{chi2}
\eeq
where the following designations are used: $x_1, x_2$ are the fractions of momentum of 
 meson carried by quark and antiquark 
correspondingly, $\xi = x_1 - x_2$, $p,~\epsilon$ are the momentum and polarizations of $P$-wave mesons.
For the mesons $\chi_1, h$ the polarization $\epsilon$ is described by the four vector $\epsilon_{\mu}$,
for the $\chi_2$ meson the polarization $\epsilon$ is described by the tensor $\epsilon_{\mu \nu}$.
The factor $[z,-z]$, that 
makes matrix elements (\ref{chi0})-(\ref{chi2})
gauge invariant, is defined as 
\beq
[z, -z] = P \exp[i g \int_{-z}^z d x^{\mu} A_{\mu} (x) ].
\eeq
In applications it is useful to rewrite the four-vector $\rho_{\mu}$ in the following way.
Evidently, one can write the polarization of the $\chi_2$ meson in terms of the polarization of 
two vector mesons. Thus for the 
transverse polarization of the meson $\chi_2$ one has 
$\epsilon^{\mu \nu}_{\lambda=\pm 1} = (\epsilon_{\lambda=\pm1}^{\mu}\cdot \epsilon_{\lambda=0}^{\nu} +
\epsilon_{\lambda=0}^{\nu}\cdot \epsilon_{\lambda=\pm1}^{\mu})/\sqrt 2$~ 
($\epsilon^+_{\mu \nu} \epsilon^{\mu \nu}=1$). 
If we further contract the polarization tensor $\epsilon_{\mu \nu}$ with lightlike four-vector $z$, 
to the leading twist accuracy we will get 
$\epsilon^{\mu \nu} z_{\nu} = \epsilon_{\lambda=\pm1}^{\mu} (pz)/(\sqrt 2 M_{\chi_2})$ or 
$\rho^{\mu}=\epsilon_{\lambda=\pm1}^{\mu}/(\sqrt 2 M_{\chi_2})$. This form of the vector 
$\rho$ can be used in the calculation with the leading twist accuracy. It should be noted that 
the states of the $\chi_2$ meson with the polarizations $\lambda=\pm 2$ give
contribution only to higher twist DAs. Since, this paper is devoted to the study of the 
leading twist DAs we don't consider these states.

The functions without primes $\phi_{\chi0}(\xi), \phi_{\chi1}(\xi), \phi_{h}(\xi), 
\phi_{\chi2}(\xi), \tilde {\phi}_{\chi2}(\xi)$ are $\xi$ odd and they are normalized as 
\beq
\int_{-1}^{1}d \xi~ \xi \phi (\xi) = 1. 
\label{norm1}
\eeq
The functions with primes $\phi'_{\chi1}(\xi), \phi'_{h}(\xi)$ are $\xi$ even 
and they are normalized as 
\beq
\int_{-1}^{1}d \xi~  \phi' (\xi) = 1. 
\eeq
In this paper all DAs will be parameterized by their moments
\beq
\langle \xi^n \rangle=\int_{-1}^1 d \xi~ \xi^n \phi(\xi).
\label{def}
\eeq
Evidently, for the DAs with primes all odd moments are zero. For the DAs without primes 
all even moments are zero. To separate the moments of different DAs of the 
$\chi_2$ meson, below we are going to use the following designations: $\langle \xi^n \rangle$
for the moments of the function $\phi_{\chi_2}(\xi)$, $\langle \tilde{\xi}^n \rangle$
for the moments of the function $\tilde {\phi}_{\chi_2}(\xi)$.

The distribution amplitudes $\phi(\xi)$ and the constants $f$ that parameterizes 
corresponding currents (\ref{chi0})-(\ref{chi2}) are scale dependent objects. 
For applications it is useful to write how they depend on the scale \cite{Chernyak:1983ej}.
To do this we expand DA in the series 
\begin{eqnarray}
\phi (\xi, \mu) = \frac 3 4 (1 - \xi^2) \biggl [  \sum_{n=0}^{\infty} a_n(\mu) C_n^{3/2} ( \xi ) \biggr ],
\label{conf_exp}
\end{eqnarray}
where $C_n^{3/2}( \xi)$ are Gegenbauer polynomials. At the leading logarithmic accuracy the 
coefficients $a_n$ are renormalized multiplicatively 
\begin{eqnarray}
a_n(\mu) = \biggl ( \frac {\alpha_s( \mu)} {\alpha_s( \mu_0)} \biggr )^{ \gamma_n / {b_0}} a_n^{L, T}(\mu_0),
\label{ren}
\end{eqnarray}
where $\gamma_n$ are the anomalous dimensions. For the current $\bar Q \sigma_{\mu \nu} [z,-z]Q$ the anomalous 
dimentions are 
\beq
\gamma_n &=& C_f \biggl (1 + 4 \sum_{j=2}^{n+1} \frac 1 j \biggr ), ~~ b_0= 11 - \frac 2 3 n_{\rm fl},~~ C_f= \frac 4 3,
\label{an_dimT}
\eeq
for the other currents 
\beq
\gamma_n &=& C_f \biggl ( 1- \frac 2 {(n+1) (n+2)} + 4 \sum_{j=2}^{n+1} \frac 1 j \biggr ). \nonumber \\
\label{an_dimVA}
\eeq
It is clear that the DAs without primes contain only n-odd  terms in series (\ref{conf_exp}). DAs 
with primes contain n-even terms in the series. 

The constants defined in equations (\ref{chi0})-(\ref{chi2}) are multiplicatively renormalizable. 
Using formulas (\ref{conf_exp})-(\ref{an_dimVA}) one can determine the evolution of these constants
\beq
f (\mu) = \biggl ( \frac {\alpha_s( \mu)} {\alpha_s( \mu_0)} \biggr )^{ \gamma / {b_0}} f(\mu_0). 
\eeq
For the constants $f_{\chi_0}, f_h, f_{\chi_2}$ the anomalous dimentions $\gamma$ is equal 
to $8/3 C_f$, for the constants $f_{\chi_1}, \tilde {f}_{\chi_2}$ $\gamma=3 C_f$, for the constant $f'_h$ $\gamma=C_f$, 
for the constant $f'_{\chi_1}$ $\gamma=0$. It should be noted that if the anomalous 
dimensions $\gamma$ of the constants $f$ are factored from sum (\ref{conf_exp}), the anomalous dimensions
of the remaining terms equal to the difference $\gamma_n-\gamma$.

\section{The distribution amplitudes at the leading order approximation in relative velocity.}

In this section the DAs under study will be considered at the 
leading order approximation in relative velocity of quark-antiquark pairs inside 
the mesons. First let us consider the DA  of  the $\chi_0$ meson. The moments of this DA can be 
represented as follows
\beq
f_{\chi_0} (pz)^{n+1} < \xi^n >_{\chi_0} =
\langle \chi_0 | \bar Q \hat z (-i z {\overset {\leftrightarrow} {D}} )^n Q | 0 \rangle. 
\label{mtr}
\eeq
To get the expressions for the constant $f_{\chi_0}$ and the moment $<\xi^n>_{\chi_0}$ one needs 
to calculate the matrix element in the right hand side. At the leading order approximation 
this calculation can be done using projector \cite{Bodwin:2002hg, Braaten:2002fi}
\beq
\bar Q (\bar p ) Q( p ) \to \int dq \frac {\varphi (-q^2)} {\sqrt {3 m_Q}} 
\frac 1 {4 \sqrt 2 E (E+m_Q)} (\hat {\bar p} - m_Q ) \Gamma (\hat P + 2 E) (\hat p +m_Q), 
\label{prj}
\eeq
where $P,q$ are the total and relative momentum of the $Q \bar Q$ pair, $m_Q$ is the 
mass of the quark $Q$, $p=P/2+q, \bar p =P/2-q$, 
$E^2=P^2/4=m_Q^2-q^2$. The matrix $\Gamma=\gamma_5, \hat {e}_S$ for the spin singlet and spin triplet 
quark-antiquark pair correspondingly, where $e_S$ is the spin polarization of this pair. 
The scalar products 
$P \cdot e_S=0, P \cdot q=0$. In the center mass frame the $dq$ is reduced to  the $d^3 {\bf q}/(2 \pi)^3$ 
and the function $\varphi (-q^2)$ is reduced to the usual nonrelativistic wave function $\phi (\bf {q^2})$. 
For the $\chi_0$ meson the wave function $\varphi (\bf {q^2})$ can be written in the form 
\beq
\varphi ({\bf {q^2}}) = \frac {{\bf e_S \cdot q}} {\sqrt 3} \psi ({\bf q}).
\label{wf1}
\eeq
At the leading order approximation in relative velocity the function $\psi(\bf q)$ is universal function for all 
$P$-wave mesons. It is normalized as 
\beq
\int \frac {d^3 {\bf q}} {(2 \pi)^3} {\bf q^i} {\bf q^j} |\psi({\bf q})|^2 = \delta^{ij}.
\eeq
With this normalization of the function $\psi({\bf q})$, the function $\varphi({\bf q})$ is normalized as
\beq
\int \frac {d^3 {\bf q}} {(2 \pi)^3} | \varphi (\bf {q^2})|^2=1.
\eeq
The same normalization condition will be used for the wave functions of all mesons 
under consideration. Using equations (\ref{mtr}), (\ref{prj}) and (\ref{wf1}) 
one gets the result
\beq
f_{\chi_0} <\xi^{n+1}>_{\chi_0} = -2^{n+1}  \sqrt{ \frac 2 m_Q} \frac {A_{n+2}} {M^{n+1}_{\chi_0}} \frac 1 {n+3},
\label{Mchi0}
\eeq
where $M_{\chi_0}$ is the mass of the $\chi_0$ meson, $A_{n}$ equals to 
\beq
A_{n} = \int \frac {d^3 {\bf q}} {(2 \pi)^3} {|\bf q|}^n \psi({\bf q}). 
\eeq
It should be noted that because of the  Coulombic part of the nonrelativistic QCD
potential the right hand side of equation (\ref{Mchi0}) is ultraviolet divergent 
\cite{Bodwin:1994jh, Bodwin:2006dn}. 
The moments of the DA in the left hand side are QCD operators (see equation (\ref{mtr}) ). 
In full QCD the Coulombic part of the nonrelativistic  potential corresponds to the 
rescattering of the quark-antiquark pair of the QCD operators which is 
also ultraviolet divergent \cite{Chernyak:1983ej}. To control the 
divergences in the right and left hand sides of equation (\ref{Mchi0})  
it is assumed that both sides are regularized within dimensional regularization.

It is interesting to note that relation (\ref{Mchi0}) is closely 
connected with Brodsky-Huang-Lepage (BHL) \cite{Brodsky:1981jv} procedure. 
This fact can be seen as follows. Let us rewrite this relation 
as follows:
\beq
\int d \xi \xi^{n+1} \phi_{\chi_0} (\xi) \sim \frac 1 {n+3} \frac {A_{n+2}} {m_Q^{n+1}} 
\sim \frac 1 {n+3} \int {\bf q^2} d{\bf q} \frac {{\bf q}^{n+2}}  {m_Q^{n+1}} \psi ({\bf q})
\sim \int d^3{\bf q} \biggl ( \frac {q_z }  {m_Q} \biggr )^{n+1} q_z \psi ({\bf q}).
\eeq
Note that $q_z \psi ({\bf q})$ is the $L_z=0$ component of the wave function $\varphi ({\bf q^2})$
which is the only component important for the leading twist DA. So, 
the last relation can be written as follows
\beq
\int d \xi \xi^{n+1} \phi_{\chi_0} (\xi) 
\sim \int d q_z \biggl ( \frac {q_z }  {m_Q} \biggr )^{n+1} \int d^2 {\bf q}_{\perp} \varphi_{L_z=0}({\bf q^2}).
\label{prel}
\eeq
Further, we change the variables in the right side of this equation 
\beq
{\bf q}_{\perp} \to {\bf q}_{\perp}, \quad q_z \to \xi {M_0}, \quad M_0^2 = \frac {M_Q^2 + {\bf q}_{\perp}^2 } {1-\xi^2}.
\eeq
Note also that $\xi \ll 1, {\bf q}_{\perp} \ll M_Q$ and at the leading order approximation in relative velocity of the quark-antiquark 
pair in the meson relation (\ref{prel}) can be written as follows
\beq
\int d \xi \xi^{n+1} \times \phi_{\chi_0} (\xi) 
\sim \int d \xi \xi^{n+1} \times \int d^2 {\bf q}_{\perp}~ \varphi_{L_z=0} \biggl (  \frac {M_Q^2 \xi^2 + {\bf q}_{\perp}^2 } {1-\xi^2}  \biggr ).
\eeq
So, at this level of accuracy the DA is just 
\beq
\phi_{\chi_0} (\xi)  \sim \int d^2 {\bf q}_{\perp}~ \varphi_{L_z=0} \biggl (  \frac {M_Q^2 \xi^2 + {\bf q}_{\perp}^2 } {1-\xi^2}  \biggr ),
\eeq
what coincides with  BHL procedure.

It is not difficult to get the relations for the other DAs and mesons
\beq
{\chi_1~ \mbox {meson:}} ~~~~  \phi ({\bf {q^2}}) &=& 
\frac { \epsilon_{ijk} { {\bf e}^i {\bf e}_S^j {\bf q}^k }} {\sqrt 2} \psi ({\bf q}), \nonumber \\
{h~ \mbox {meson:}}~~~~  \phi ({\bf {q^2}}) &=& 
\frac { {\bf e \cdot  q  } } {\sqrt 2} \psi ({\bf q}), \nonumber \\
{\chi_2~ \mbox {meson:}}~~~~  \phi ({\bf {q^2}}) &=& 
 {\bf e}_{ij} \bf {e_S}^i   {\bf q}^j   \psi ({\bf q}),
\eeq
where $\bf e$ is the polarization three vector of the $\chi_1$ and $h$ mesons,
${\bf e}_{ij}$ is the polarization tensor of the $\chi_2$ meson. Using these 
expressions one can get 
\beq
f_{\chi_1} <\xi^{n+1}>_{\chi_1} = -2^{n+1}  \sqrt{ \frac 3 {m_Q}} \frac {A_{n+2}} {M^{n+1}_{\chi_1}} \frac 1 {n+3}, \nonumber \\
f'_{\chi_1} <\xi^n>'_{\chi_1} =2^{n+1} i  \sqrt{ \frac {12} {m_Q} } \frac {A_{n+2}} {M^{n+1}_{\chi_1}} \frac 1 {(n+1)(n+3)}, \nonumber \\
f_{h} <\xi^{n+1}>_{h} = -2^{n+1}  \sqrt{ \frac 6 {m_Q}} \frac {A_{n+2}} {M^{n+1}_{h}} \frac 1 {n+3}, \nonumber \\
f'_{h} <\xi^n>'_{h} = 2^{n+1} i  \sqrt{ \frac 6 {m_Q} } \frac {A_{n+2}} {M^{n+1}_{h}} \frac 1 {(n+1)(n+3)}, \nonumber \\
f_{\chi_2} <\xi^{n+1}>_{\chi_2} = 
-2^{n+1}  \sqrt{ \frac 4 {m_Q}}  \frac {A_{n+2}} {M^{n+1}_{\chi_2}} \frac 1 {n+3}, \nonumber \\
\tilde {f}_{\chi_2} <\tilde {\xi}^{n+1}>_{\chi_2} = - 2^{n+1}  \sqrt{ \frac {6} {m_Q} } \frac {A_{n+2}} {M^{n+1}_{\chi_2}} \frac 1 {n+3}.
\label{rel}
\eeq
The moments with primes correspond to the DAs with primes. As it was noted above,
since all the DAs without primes are $\xi$-odd only odd moments 
of these DAs are nonzero. Similarly, for the DAs with primes only even moments 
different from zero. For these reason we took $(n+1)$th moments of DAs without 
primes and $n$th moments with primes, where n is assumed to be even. 
It should be noted that since the function $\psi ({\bf q})$ is universal for the $P$-wave mesons, 
the $A_{n}$ is the same for all mesons under study.  From relation (\ref{rel}) one can conclude that all 
the constants $f_i$ can be expressed through the only constant which will be designated as $F$
\beq
F={\sqrt 3} {f_{\chi_0}} = {\sqrt 2 } {f_{\chi_1}} = i \frac {f'_{\chi_1}} {\sqrt 2} = 
f_{h}= i f'_{h} = \sqrt {\frac 3 2} f_{\chi_2}=  { \tilde {f}_{\chi_2} } .
\label{const1}
\eeq
From equations (\ref{rel}) one
can also find that at the leading order approximation the $n$th moments of all functions 
without primes coincide. This means that these DAs are equal to one universal DA which will be designated as $\Phi(\xi)$
\beq
\Phi(\xi) = \phi_{\chi_0} (\xi) =  \phi_{\chi_1} (\xi)= \phi_{h} (\xi)= \phi_{\chi_2} (\xi)= \tilde {\phi}_{\chi_2} (\xi). 
\label{func1}
\eeq
The same is true for all DAs with primes, which are equal to one function which will be designated below 
as $\Psi(\xi)$
\beq
\Psi(\xi) =   \phi'_{\chi_1} (\xi)= \phi'_{h} (\xi). 
\label{func2}
\eeq
In addition, one can relate the moments of the $\Phi(\xi)$ to the moments of the $\Psi(\xi)$ as follows
\beq
<\xi^n>_{\Psi}= \frac {<\xi^{n+1}>_{\Phi}} {n+1}.
\label{mom}
\eeq
It should be noted that the constants and DAs in relations (\ref{const1})-(\ref{mom}) depend 
on scale in a different way. This means that relations (\ref{const1})-(\ref{mom}), which are valid 
at not to large scale, will be violated at sufficiently large scale. 

Recursive relation (\ref{mom}) determines the function $\Psi(\xi)$
through the function $\Phi(\xi)$. One can  guess the solution
of this relation:
\beq
\Psi(\xi) =- \int_{-1}^{\xi} dt~ \Phi(t).
\label{funcc}
\eeq
To prove that (\ref{funcc}) is the solution of relations (\ref{mom})  
one should put this function to the definition of the n-th moment (\ref{def}) and integrate the resulting 
expression by parts. {\it It is seen from  equations (\ref{func1}), (\ref{func2}) and (\ref{funcc}) 
that all DAs of the $P$-wave mesons are defined through the universal function $\Psi(\xi)$.} 
It should be noted that this fact results from the nonrelativistic spin-symmetry, 
which holds at leading order in the heavy-quark velocity. Below equations 
(\ref{func1}) and (\ref{funcc}) will be used to build the models for 
the function $\Phi(\xi)$ and $\Psi(\xi)$ of $P$-wave charmonia.

At the end of this section it is interesting to discuss the question about 
relativistic corrections to the matrix elements involving $P$-wave quarkonia.
If we ignore the contribution coming from the higher Fock states, relativistic 
corrections to the matrix involving, for instance, the $\chi_0$ meson is given by the 
matrix element 
\beq
\langle v^n \rangle_{P} = \frac 1 {(m_Q^*)^2}
\frac { \langle \chi_0 | \chi^+ \bigl ({\bf \sigma D } \bigr )  { \bf D^2}  \psi | 0 \rangle }
{ \langle \chi_0 | \chi^+  \bigl ( {\bf \sigma D } \bigr )    \psi | 0 \rangle },
\eeq
where $\psi$ and $\chi^+$ are Pauli spinor fields that annihilate a 
quark and an antiquark respectively, ${\bf \sigma }$ are Pauli matrixes, 
$m_Q^*$ is the quark pole mass. Using equation (\ref{Mchi0}) 
it is not difficult to obtain general formula that connects 
the moment $\langle \xi^{n+1} \rangle$ with the matrix element $\langle v^n \rangle_{P}$
at the leading order approximation in relative velocity
\beq
\langle v^n \rangle_{P} = \frac {A_{n+2}} {A_{2}} + O(v^{n+2}) = 
\frac {n+3} {3} \langle \xi^{n+1} \rangle + O(v^{n+2}).
\label{rel1}
\eeq
Although the derivation was done for the $\chi_0$ meson,
at the leading order approximation in relative velocity the matrix 
element $\langle v^n \rangle_{P}$ is universal for all $P$-wave mesons. 

\section{Charmonia distribution amplitudes within QCD sum rules.} 

  In this section QCD sum rules approach \cite{Shifman:1978bx, Shifman:1978by} will be applied to the calculation 
of the moments of the $P$-wave charmonia DAs. The problem which can be met if one tries to apply 
QCD sum rules to the calculation of these moments  is that it is not 
possible to write QCD sum rules for one DA. Commonly, the contribution of different mesons and different DAs 
mix in  QCD sum rules, what does not allow us to calculate the constants and 
the moments of DAs (\ref{chi0})-(\ref{chi2}) separately. Only some combinations of different constants and moments can be 
extracted from QCD sum rules. For instance, two point QCD sum rules for the currents 
$ \bar Q(x) \gamma_{\mu} (z {\overset {\leftrightarrow} {D}}) Q(x) \cdot \bar Q(0) \gamma_{\mu} 
(z {\overset {\leftrightarrow} {D}})^{n+1} Q(0) $  can be written as follows
\beq
\frac {f_{\chi_{c0}}^2 \langle \xi^{n+1} \rangle_{\chi_{c0}} }  { (m_{\chi_{c0}}^2+Q^2)^{m+1} } +
\frac {f_{\chi_{c2}}^2 \langle \xi^{n+1} \rangle_{\chi_{c2}} }  { (m_{\chi_{c2}}^2+Q^2)^{m+1} }= 
\frac 1 {\pi} \int_{4 m_c^2}^{s_0} ds ~ \frac {\mbox{Im}~ \Pi_{\rm pert}(s,n)} {(s+Q^2)^{m+1} }  + \Pi^{(m)}_{\rm npert}(Q^2,n).
\label{sm1}
\eeq
The expressions for the $\mbox {Im}~ \Pi_{\rm pert} (s,n)$, $\Pi^{(m)}_{\rm npert}(Q^2,n)$ 
will be given below (equations (\ref{pert}),(\ref{power})). It is seen from this example 
that it is not possible to extract the constants $f_{\chi_{c0}}, f_{\chi_{c2}}$ or the moments
$\langle \xi^{n+1} \rangle_{\chi_{c0}}, \langle \xi^{n+1} \rangle_{\chi_{c2}}$ from (\ref{sm1})  separately. 
Evidently, this strongly restricts the accuracy of the calculation. The only QCD sum rules 
which are free from this problem  are two point sum rules for the currents
$ \bar Q(x) \gamma_{\mu} \gamma_5 (z {\overset {\leftrightarrow} {D}}) Q(x) \cdot \bar Q(0) \gamma_{\mu} 
\gamma_5 (z {\overset {\leftrightarrow} {D}})^{n+1} Q(0) $
\beq
\frac { f_{h_c}^2 \langle \xi^{n+1} \rangle_{h_c} }  { (m_{h_c}^2+Q^2)^{m+1} } = 
\frac 1 {\pi} \int_{4 m_c^2}^{s_0} ds ~ \frac {\mbox{Im}~ \Pi_{\rm pert}(s,n)} {(s+Q^2)^{m+1} }  + \Pi^{(m)}_{\rm npert}(Q^2,n),
\label{sm2}
\eeq
where $\mbox {Im}~ \Pi_{\rm pert} (s,n)$, $\Pi^{(m)}_{\rm npert}(Q^2,n)$ can be written as 
\beq
\mbox {Im}~ \Pi_{\rm pert} (s,n) = \frac {3} {8 \pi} v^{n+3}  (\frac 1 {n+3} - \frac {v^2} {n+5} ), ~~~~ v^2 = 1 - \frac {4 m_c^2} {s},
\label{pert}
\eeq
\beq
\label{power}
\Pi^{(m)}_{\rm npert}(Q^2,n) &=& \Pi_1^{(m)} (Q^2,n) + \Pi_2^{(m)} (Q^2,n)+\Pi_3^{(m)} (Q^2,n), \\ \nonumber
\Pi_1^{(m)} (Q^2,n) &=& \frac {\langle \alpha_s G^2 \rangle} {24 \pi }
(m+1) \int_{-1}^1 d \xi~  \biggl (  \xi^{n+2} + \frac {n (n+1)} 4 \xi^{n} (1- \xi^2) \biggr )
\frac {(1- \xi^2)^{m+2}} {\bigl ( 4 m_c^2  + Q^2 (1- \xi^2 ) \bigr )^{m+2} }, \\ \nonumber
\Pi_2^{(m)} (Q^2,n) &=& - \frac {\langle \alpha_s G^2 \rangle} {6 \pi} m_c^2 ( m^2+ 3m +2)
\int_{-1}^1 d \xi~ 
  \xi^{n+2} \bigl ( 1+ 3 \xi^2 \bigr ) \frac {(1- \xi^2)^{m+1}} {\bigl ( 4 m_c^2  + Q^2 (1- \xi^2 ) \bigr )^{m+3} }, \\ \nonumber
\Pi_3^{(m)} (Q^2,n) &=&  \frac {\langle \alpha_s G^2 \rangle} {384 \pi }
n(n+1) (m+1) \int_{-1}^1 d \xi~  \xi^{n} 
\frac {(1- \xi^2)^{m+3}} {\bigl ( 4 m_c^2  + Q^2 (1- \xi^2 ) \bigr )^{m+2} }.
\eeq
In the calculation we take $Q^2=4 m_c^2$ \cite{Reinders:1984sr}. 
In the numerical analysis of QCD sum rules the values of parameters $m_c$ and 
$\langle \alpha_s G^2/ \pi \rangle$ will be taken from paper \cite{Reinders:1984sr}:
\beq
m_c = 1.24 \pm 0.02 ~\mbox {GeV}, ~~ \langle \frac {\alpha_s} {\pi} G^2 \rangle = 0.012 \pm 30 \% ~\mbox {GeV}^4.
\label{param}
\eeq
First sum rules (\ref{sm2}) will be applied to the calculation of the constant $f_{h_c}^2$. 
It is not difficult to express the constant $f_{h_c}^2$  from equation (\ref{sm2}) at $n=0$ as a function
of $m$. For too small values of $m$ ($m<m_1$) there is large contributions from  higher resonances 
and continuum which spoil sum rules (\ref{sm2}). Although for $m \gg m_1$ these contributions
are strongly suppressed, it is not possible to apply sum rules for too large $m$ ($m>m_2$) 
since the contribution arising from higher dimensional vacuum condensates rapidly grows with $m$ what invalidates 
our approximation. If $m_1<m_2$ there is some region of applicability of sum rules (\ref{sm2}) 
$[m_1, m_2]$ where the resonance and the higher dimensional vacuum condensates contributions 
are not too large. Within this region $f_{h_c}^2$ as a function of $m$ varies slowly and one can 
determine the value of this constant. The value of the continuum threshold $s_0$ must be 
taken so that to appear stability region $[m_1, m_2]$. Our calculation shows that for 
the central values of  parameters (\ref{param}) there exists stability region for 
$s_0>(4.3$~GeV$)^2$. If the value of the continuum threshold $s_0$ is varied in the 
region $s_0 \in (4.3^2,\infty)$~GeV$^2$, the value of the constant $f_{h_c}^2$ can be written as 
$f_{h_c}^2=(0.037 \pm 0.005)$~GeV$^2$.
In addition to the uncertainty due to the variation of the value of $s_0$, there are uncertainties
due to the variation of the values of the $m_c$ (which is $\pm 0.004$) and 
$\langle  {\alpha_s}/ {\pi} G^2 \rangle$ (which is $\pm 0.001$). The last
source of uncertainty is the radiative corrections to the perturbative density $\mbox {Im} \Pi_{\rm pert} (s,n)$, 
which will be estimated as $\alpha_s (m_c)/\pi \sim 13 \%$.
Adding these uncertainties in quadrature, one gets
\beq
f_{h_c}^2 (\mu \sim m_c)=( 0.037 \pm  0.007) \mbox{GeV}^2.
\label{constf}
\eeq
As it was noted above, the value of the constant $f_{h_c}^2$ is scale dependent 
quantity. The characteristic scale of QCD sum rules is $\sim m_c$. 
This means that the value of the constant $f_{h_c}^2$ is determined 
at the scale $\sim m_c$, as it is shown in (\ref{constf}).

Next let us consider the moments $\langle \xi^3 \rangle_{h_c}, \langle \xi^5 \rangle_{h_c}, 
\langle \xi^7 \rangle_{h_c}$.  
However, instead of considering QCD sum rules for $n=2,4,6$ we will consider the ratios of sum rules at 
$n=2,4,6$ and the sum rules at $n=0$. Such approach improves the accuracy of the calculation (see paper 
\cite{Braguta:2006wr} for details). 
The analisys similar to that for the constant $f_{h_c}^2$ gives 
\beq
\langle \xi^3 \rangle &=& 0.18 \pm 0.03 , \nonumber \\
\langle \xi^5 \rangle &=& 0.050 \pm 0.010, \nonumber \\
\langle \xi^7 \rangle &=& 0.017 \pm 0.004.
\label{res}
\eeq
It should be noted that these moments are defined at the scale $\sim m_c$.
It should be also noted that the values of the moments (\ref{res}) are in good agreement 
with potential model estimation (see paper \cite{Braguta:2006wr}).
For instance, within Buchmuller-Tye potential model \cite{Buchmuller:1980su} $\langle \xi^3 \rangle=0.18,
\langle \xi^5 \rangle=0.047, \langle \xi^7 \rangle=0.016$;
within Cornell potential model \cite{Eichten:1978tg} $\langle \xi^3 \rangle=0.16,
\langle \xi^5 \rangle=0.040, \langle \xi^7 \rangle=0.013$.

Using values (\ref{res}) one can find the matrix elements 
that control relativisitic corrections to any process with $P$-wave 
charmonia in the initial or final state. The relationships between 
these matrix elements and the moments are given in equation (\ref{rel1}). 
These relations are valid up to the higher order relativistic corrections, 
which can be estimated as $\langle v^2 \rangle$. Taking into account this 
additional source of uncertainty one gets
\beq
\langle v^2 \rangle_P &=& \frac 5 3 {\langle \xi^3 \rangle_P}  = 0.30 \pm 0.10 , \nonumber \\
\langle v^4 \rangle_P &=& \frac 7 3 {\langle \xi^5 \rangle_P}  = 0.12 \pm 0.04, \nonumber \\
\langle v^6 \rangle_P &=& \frac 9 3 {\langle \xi^7 \rangle_P}  = 0.051 \pm 0.018.
\eeq
In the next sections  results (\ref{res}) will be used to build the 
model for the DAs $\Phi(\xi)$ and $\Psi(\xi)$.

\section{Model for charmonia distribution amplitudes.}

To build the model of the function $\Phi(\xi,\mu \sim m_c)$ we use 
Borel version \cite{Shifman:1978bx, Shifman:1978by} of sum rules (\ref{sm2}) 
but without continuum contribution and power corrections 
\beq
f_{h_c}^2 \langle \xi^{n+1} \rangle e^{-  {m_{h_c}^2}/ {M^2}} = 
\frac {M^2} {4 \pi^2} \int_{-1}^1 d \xi ~\xi^{n+2} ~\frac 3 4 (1- \xi^2) \mbox{exp} \biggl (  - \frac {4 m_c^2} {M^2} \frac 1 {1-\xi^2} \biggr ).
\label{snm}
\eeq
Evidently, within this approximation the function $\Phi(\xi,\mu \sim m_c)$ can be written in the form
\beq
\Phi(\xi, \mu \sim m_c) = c( \beta_P ) (1- \xi^2) ~\xi~ \mbox{exp} \biggl (  -  \frac {\beta_P} {1-\xi^2} \biggr ),
\label{model}
\eeq
where $ c (\beta_P)$ is a normalization constant and $\beta_P$ is some constant.
We propose function (\ref{model}) as the model for DAs $\Phi(\xi, \mu \sim m_c)$. 
To fix the constant $\beta_P$ the value of the moment $\langle \xi^2 \rangle$ (\ref{res}) will be used. 
Thus we get $\beta_P = 3.4^{+1.5}_{-0.9}$. The constant $c(\beta)$ can be determined 
from normalization condition (\ref{norm1}). The moments of the function (\ref{model})
are
\beq
\langle \xi^3 \rangle &=& ~0.18 \pm 0.03,  \\
\langle \xi^5 \rangle &=& ~0.047  \pm 0.014, \nonumber \\
\langle \xi^7 \rangle &=& 0.015 \pm 0.006. \nonumber  
\eeq
Using the model for the function $\Phi(\xi, \mu \sim m_c)$ and equation 
(\ref{funcc}) one can get the model of the DA $\Psi(\xi, \mu \sim m_c)$
\beq
\Psi(\xi, \mu \sim m_c) =- \int_{-1}^{\xi} dt \Phi(t, \mu \sim m_c)= \frac {c(\beta_P)} 2 (1-\xi^2)^2~ E_3\biggl ( \frac {\beta_P} {1-\xi^2}  \biggr ) ,
\label{func3}
\eeq
where the function $E_3(z)$ is the exponential integral function
\beq
E_3(z)=\int_1^{\infty} dt \frac {e^{-z t}} {t^3}.
\eeq
To determine DAs (\ref{chi0})-(\ref{chi2}) at a scale different from $m_c$ 
one should use relations (\ref{func1}), (\ref{func2}) and than evolution 
equations for the corresponding DAs. 

\section{Conclusion.}

In this paper we have considered the leading twist distribution amplitudes (DA)
of $P$-wave nonrelativistic mesons. At the leading order approximation 
in relative velocity of quark-antiquark pair inside the mesons these 
functions can be expressed through one universal DA. We have derived 
the relations between the DAs of $P$-wave nonrelativistic mesons and 
the universal DA. 

As an example, we have considered the DAs of the $P$-wave charmonia mesons. 
Within QCD sum rules we found the moments of the leading twist DA of $h_c$
meson, what allowed us to build the model of the universal DA for the $P$-wave 
charmonia mesons. 

In addition, we have found the relations between the moments and the nonrelativistic QCD matrix elements 
that control relativistic corrections to any amplitude involving $P$-wave charmonia. 
The calculation shows that characteristic size of these corrections is $\sim 30 \%$. 

This work was partially supported by Russian Foundation of Basic Research under grant 07-02-00417.
The work of V. Braguta was partially supported by  CRDF grant Y3-P-11-05 and president grant MK-2996.2007.2.
The work of A. Luchinsky was partially supported by  president grant MK-110.2008.2 and Russian Science Support
Foundation.

\end{document}